\documentclass[preprint,12pt]{elsarticle}
\usepackage[hyphens]{url}
\RequirePackage[OT1]{fontenc}
\RequirePackage{natbib}
\RequirePackage[colorlinks,citecolor=blue,urlcolor=blue,breaklinks]{hyperref}
\usepackage{amsgen,amsmath,amsthm,amstext,amsbsy,amsopn,amssymb,amscd}
\usepackage{graphicx, subfigure}
\usepackage{cases}
\usepackage{dcolumn}
\usepackage{bm}
\usepackage{booktabs}
\usepackage[T1]{fontenc}
\usepackage{natbib}
\biboptions{authoryear, round}
\usepackage{color}
\RequirePackage{hypernat}

\newtheorem{remark}{\bf Remark}[section]
\newtheorem{lemma}{\bf Lemma}[section]
\newtheorem{proposition}{\bf Proposition}[section]

\newtheorem{definition}{\bf Definition}[section]
\newtheorem{theorem}{\bf Theorem}[section]
\numberwithin{equation}{section}

\theoremstyle{plain}
\newcommand{\iid}{\stackrel{\mathrm{iid}}{\sim}}
\newcommand{\ind}{\stackrel{\mathrm{ind}}{\sim}}

\journal{Journal of Multivariate Analysis}
\begin{document}
\begin{frontmatter}
\title{Capturing the Severity of Type II Errors in High-Dimensional Multiple Testing}

\author[label1]{Li He}
\address[label1]{Merck Research Laboratories, West Point, PA 19486}
 \ead{li.he@merck.com}
\author[label2]{Sanat K, Sarkar}
\address[label2]{Department of Statistics, Temple University,
Philadelphia, PA, US, 19122}
 \ead{sanat@temple.edu}
\author[label3]{Zhigen Zhao}
\address[label3]{Department of Statistics, Temple University,
Philadelphia, PA, US, 19122}
 \ead{zhaozhg@temple.edu}

\begin{abstract}
The severity of type II errors is frequently ignored when deriving a multiple testing procedure, even though utilizing it properly can greatly help in making correct decisions. This paper puts forward a theory behind developing a multiple testing procedure that can incorporate the type II error severity and is optimal in the sense of minimizing a measure of false non-discoveries among all procedures controlling a measure of false discoveries. The theory is developed under a general model allowing arbitrary dependence by taking a compound decision theoretic approach to multiple testing with a loss function incorporating the type II error severity. We present this optimal procedure in its oracle form and offer numerical evidence of its superior performance over relevant competitors.
\end{abstract}

\begin{keyword}
Bayes rule
\sep Compound decision theory
\sep Oracle procedure
\sep Multiple testing
\sep Weighted marginal false discovery rate
\sep Weighted marginal false non-discovery rate
\end{keyword}

\end{frontmatter}

\section{Introduction}\label{sec:intro}
Simultaneous testing of multiple hypotheses is an integral part of
analyzing high-dimensional data from modern scientific
investigations like those in genomics, brain imaging, astronomy, and
many others, making multiple testing an  area of current
importance and intense statistical research. A variety of multiple
testing methods have been put forward in the literature from both
frequentist and Bayesian perspectives. However, the theories behind
the developments of these methods are mostly driven by the
overreaching goal of controlling an overall measure of type I errors
or false discoveries, with other fundamentally important statistical
issues often being ignored. For instance, in many of the
aforementioned experiments there is a cost associated with the error
of making a false discovery or missing a true discovery, and this
cost increases with increasing severity of that error. This is an
important issue not often taken into account when
developing multiple testing procedures.

A Bayesian decision theoretic approach can yield a powerful multiple testing method not only incorporating costs of false and missed
discoveries but also simultaneously addressing dependency, optimality, and multiplicity (\cite{Sun:Cai:2007,Sun:Cai:2009}). This motivates us to take a similar approach, but in a more general framework that conforms more to the present problem, that is, to address the aforementioned issue related to severity of errors. Before explaining this generalization, let us first briefly outline the approach taken in \cite{Sun:Cai:2007,Sun:Cai:2009}.

Given a set of observations ${\mathbf{X}}=(X_{1}, \ldots, X_{m}) \sim
f({\mathbf{x}}, \mbox{\boldmath$\theta$})$, where
$\mbox{\boldmath$\theta$}= (\theta_{1}, \dots, \theta_{m} ) \in
\{0,1\}^m$, consider the problem of deciding between
$H_i:\theta_i=0$ and $\bar H_i: \theta_i=1$ simultaneously for $i=1,
\ldots, m$, assuming that $X_i~|~\theta_i \stackrel {ind} \sim
(1-\theta_i)f_{0}(x_i) + \theta_i f_1(x_i)$, for some given
densities $f_0$ and $f_1$, and $\theta_i \sim Bernoulli (1- \pi_0)$. \cite{Sun:Cai:2007,Sun:Cai:2009} started with the following uniformly weighted 0-1 loss
function:  \begin {eqnarray} L_{\lambda}(\mbox{\boldmath$\delta$(\bf
{X})}, \mbox{\boldmath$\theta$}) = \frac{1}{m} \sum_{i=1}^m \left
\{\lambda (1 - \theta_i) \delta_i(\mathbf{X}) + \theta_i (1 -
\delta_i(\mathbf{X})) \right \}, \end {eqnarray} for a decision rule $\mbox{\boldmath$\delta$}(\mathbf{X}) =
(\delta_{1}(\mathbf{X}), \dots, \delta_{m}(\mathbf{X}) ) \in
\{0,1\}^m$, where $\lambda$ is the
relative cost of making a false discovery (type I error) to that of
missing a true discovery (type II error) and assumed to be constant over all the hypotheses. They considered the Bayes rule associated with this loss function and showed that it is also optimal from a multiple testing point of view. Specifically, given any $\alpha \in (0,1)$, there exists a $\lambda \equiv \lambda (\alpha)$ for which it controls the marginal false discovery rate,
\begin {eqnarray} {\rm mFDR} = \frac{ E \left [ \sum_{i=1}^m \delta_i
(\mathbf{X}) (1 - \theta_i)\right ] }{E  \left [\sum_{i=1}^m\delta_i
(\mathbf{X}) \right] },  \nonumber \end {eqnarray}
at $\alpha$, and minimizes the marginal false non-discovery rate,
\begin {eqnarray} {\rm mFNR}= \frac{ E \left [ \sum_{i=1}^m \{1-\delta_i(\mathbf{X}) \} \theta_i \right ]}{ E  \left [\sum_{i=1}^m \left \{1 - \delta_i(\mathbf{X}) \right \} \right] }, \nonumber \end {eqnarray}
among all decision rules defined in terms of statistics satisfying a monotone likelihood ratio condition (MLR) and controlling the mFDR at $\alpha$.
They expressed this optimal procedure in an alternative form using hypothesis specific test statistics defined in terms of the local FDR measure [lfdr, \citet{Efron:book:2010}], and called it the oracle procedure.
%They constructed a data-driven version of it under independence \citep{Sun:Cai:2007} or Markov dependence \citep{Sun:Cai:2009} for the $\theta_i$'s.
They provided numerical evidence showing that their oracle procedure can
outperform its competitors, such as those in
\citet{Benjamini:Hochberg:1995} and \citet{Genovese:Wasserman:2002}.
%and theoretically proved that their data-driven procedure is asymptotically equivalent to the corresponding oracle procedure. Later, \citet{Xie:Cai:Maris:Li:2011} showed that the results in \citet{Sun:Cai:2007} continue to hold under the so-called short-range dependence among the test statistics.

Clearly, the loss function used in the above formulation is somewhat
simplistic. It gives equal importance to all type I errors as well
as to all type II errors. While it might be reasonable to treat the
type I errors equally in terms of severity and attach a fixed cost
to all of them, it is often unrealistic to do so for type II errors.
For instance, in a microarray experiment, there might be a fixed
cost of doing a targeted experiment to verify that each gene is
active and the loss due to making a false discovery might be that
cost (which is being wasted in case the gene is found to be
inactive). However, it would be unrealistic to assume that the loss
in identifying a truly active gene as inactive does not depend on
how strong is the expected signal that has been missed. In fact, it
might reasonably be proportional to the difference
\citep{Duncan:1965, Waller:Duncan:1969, Scott:Berger:2005} or even
to the squared difference between the expected values of the missed
and no signals.

In other words, the above formulation needs to be generalized
conforming it more to the reality in modern high-dimensional multiple
testing. With that in mind, we consider testing $H_i: \mu_i =
\mu_{i0}$ against its one or two-sided alternative, for some
specified values $\mu_{i0}$, simultaneously for $i=1, \ldots, m$,
under the following model:
%$\mu_i \equiv \mu_i(\theta_i)$
\begin{eqnarray} \label{model}
 \mathbf{X} \mid
\mbox{\boldmath{$\mu$}}, \mbox{\boldmath{$\theta$}} & \sim &
f(\mathbf{x} \mid \mbox{\boldmath{$\mu$}}), \; \mbox{with} \; \mbox{\boldmath{$\mu$}} = (\mu_1, \ldots, \mu_m), \; \mbox{\boldmath{$\theta$}} = (\theta_1, \ldots, \theta_m) \nonumber \\
\mu_{i} \mid \theta_{i} & \sim & (1-\theta_{i}) I(\mu_i = \mu_{i0}) + \theta_{i} h(\mu_i - \mu_{i0}) \\
\theta_{i} & \sim & Bernoulli(1-\pi_{0}), \nonumber \end{eqnarray} given a density $h$, and under the following more general loss function:
\begin {eqnarray} \label{loss}
\\
L_{\lambda, s}(\mbox{\boldmath{$\delta$}}({\bf X}), \mbox{\boldmath{$\mu$}}, \mbox{\boldmath{$\theta$}}) = \frac{1}{m} \sum_{i=1}^m \left \{\lambda (1 - \theta_i) \delta_i({\bf X}) + s(\mu_i-\mu_{i0})\theta_i (1 - \delta_i({\bf X})) \right \}.\nonumber
\end {eqnarray}
We do not impose any dependence restriction on $\textbf{X}$, $\mbox{\boldmath{$\mu$}}$ or $\mbox{\boldmath{$\theta$}}$.
It is assumed that there is only a baseline cost $\lambda_1$ for each type I error (which, as argued above, is reasonable for a point null hypothesis). For each type II error, however, we assume that the cost is $\lambda_2$, the baseline cost, times $s(\mu_i - \mu_{i0})$, a function $s$ of $\mu_i - \mu_{i0}$ such that $s(0)=0$ and is non-decreasing as $\mu_i$ moves away from $\mu_{i0}$.  We call $s(\cdot)$ the {\it severity function} for type II errors. Through this function, a penalty is being imposed on making a type II error for each $H_i$; the larger the value of $|\mu_{i} - \mu_{i0}|$ is, the more severe this penalty is. The $\lambda$ equals $\lambda_1/\lambda_2$, the relative baseline cost of a type I error to a type II error. In other words, $\lambda/s(\mu_i - \mu_{i0})$ is the relative cost of a type I error to a type II error. The specific choice of $s(\cdot)$ will depend on how fast we want the cost of the type II error to increase as $\mu_i$ moves away from $\mu_{i0}$.

Our proposed loss function (\ref{loss}) is a non-uniformly weighted
0-1 loss function giving less and less weight to the type I error
relative to the type II error as the type II error gets more and more
severe as measured by the severity function. In this paper, we focus on deriving the theoretical form of an optimal multiple testing procedure from the Bayes rule
under this general loss function. Given a severity function $s$, this Bayes
rule provides an optimal multiple testing procedure in the sense of
minimizing a measure of non-discoveries subject to controlling a
measure of false discoveries at a specified level for a suitably
chosen $\lambda$. These measures of false discoveries and false non-discoveries are
of course different from the mFDR and mFNR, respectively, since we
now need to account for the weights or penalties attached to the
type II errors through the severity function that is not necessarily
equal to one. We define these newer error rates as weighted mFDR and weighted mFNR and establish the
aforementioned optimality result through these rates.
%We consider an approximate version of this optimal procedure through hypothesis specific test statistics, referred to as generalized local fdr's, which will be our proposed oracle procedure.
We study the performance of this oracle optimal procedure with its relevant competitors through two numerical studies. %The derivation of the corresponding data-driven procedure involves estimation of complicated functions which requires special attention and effort, and so we defer that to a separate communication.

The remainder of the paper is organized as follows. The development
of the Bayes rule under the loss function (\ref{loss}), its
characterization as an optimal multiple testing procedure in the
framework of weighted false discovery and false non-discovery
rates, and our oracle multiple testing procedure are given in the next section. In Section \ref{sec:comp:oracle}, we present the results of two numerical studies providing
evidence of this oracle procedure's superior performance over its relevant
competitors. We end the paper with some concluding remarks in Section \ref{sec:concluding}.

\section{Optimal Rules}\label{sec:decision}
Assuming that our problem is that of testing $H_i: \mu_i=0$
simultaneously for $i=1, \ldots, m$  under the model (\ref{model})
and the loss function $L_{\lambda, s}$ in (\ref{loss}), we do the
following in this section:  (i) determine the Bayes rule; (ii) show
that the Bayes rule with an appropriately chosen $\lambda$ provides
an optimal multiple testing procedure in the sense of minimizing a
measure of false non-discoveries among all rules that control a
measure of false discoveries at a specified level; and (iii) express
this optimal multiple testing procedure in terms of some test
statistics to define the oracle procedure in this paper.

\subsection {The Bayes rule} Let us first define \begin{align}\label{wi}
w_{i}(\mathbf{X})=E \left [s(\mu_{i}) \mid \theta_{i}=1, \mathbf{X} \right ],
\end{align} the average severity of type II errors conditional on the
data $\mathbf{X}$ and $\theta_i=1$. Then, we have the following:
\begin{theorem}
\label{thm1} Consider testing $H_i: \mu_i=0$ simultaneously for $i=1, \ldots, m$ under the model (\ref{model}) and the loss function (\ref{loss}). Then, the decision rule $\mbox{\boldmath{$\delta$}}^*(\mathbf{X}) =(\delta_1^*(\mathbf{X}), \ldots, \delta_m^*(\mathbf{X}))$, where
\begin{align}
 \label{Bayesrule}
   \delta^*_{i}(\mathbf{X})= \left\{
     \begin{array}{ll}
       1 &  \textrm{if} \quad P(\theta_{i}=0 \mid \mathbf{X}) < \dfrac{w_{i}(\mathbf{X})}{\lambda}P(\theta_{i}=1 \mid \mathbf{X})\\
       & \\ 0 &  \textrm{if} \quad P(\theta_{i}=0 \mid \mathbf{X}) > \dfrac{w_{i}(\mathbf{X})}{\lambda}P(\theta_{i}=1 \mid \mathbf{X})~,
     \end{array}
   \right.
\end{align}
is the Bayes rule.
\end{theorem}

\textbf{Proof.} For any rule $\mbox{\boldmath{$\delta$}}(\mathbf{X}) =  ( \delta_1(\mathbf{X}), \ldots, \delta_m(\mathbf{X}))$, we have
\begin{eqnarray}
& & E \left [ L_{\lambda, s}(\mbox{\boldmath{$\theta$}}, \mbox{\boldmath{$\mu$}}, \mbox{\boldmath{$\delta$}}(\mathbf{X}))\mid\mathbf{X} \right ]\nonumber \\
&=&\dfrac{1}{m}\sum_{i=1}^{m}\left\{\lambda \delta_{i}(\mathbf{X}) P(\theta_{i}=0\mid \mathbf{X})+[1-\delta_{i}(\mathbf{X})]E \left [ s(\mu_{i}) I(\theta_{i}=1) \mid  \mathbf{X} \right ] \right\} \nonumber\\
&=&\dfrac{1}{m}\sum_{i=1}^{m}\left\{\lambda \delta_{i}(\mathbf{X}) P(\theta_{i}=0\mid \mathbf{X})+[1-\delta_{i}(\mathbf{X})]E \left [s(\mu_{i}) \mid \theta_{i}=1,  \mathbf{X}\right]  P(\theta_{i}=1 \mid \mathbf{X} )\right\} \nonumber\\
&=&\dfrac{1}{m}\sum_{i=1}^{m}\left\{w_i(\mathbf{X})  P(\theta_{i}=1 \mid \mathbf{X}) + \delta_{i}(\mathbf{X})\left [ \lambda P(\theta_{i}=0 \mid \mathbf{X})
- w_i(\mathbf{X})  P(\theta_{i}=1 \mid \mathbf{X}) \right ]\right \}.  \nonumber
\end{eqnarray}
Since the first term is constant with respect to \mbox{\boldmath{$\delta$}}, given $\mathbf{X}$, it is clear that  $\mbox{\boldmath{$\delta$}}^*(\mathbf{X})$ in (\ref{Bayesrule}) is the rule for which this conditional expectation is the minimum among all $\mbox{\boldmath{$\delta$}}$, and hence is Bayes. \qed

\subsection{Optimal Multiple Testing Procedure}

Here we show that the aforementioned Bayes rule with an
appropriately chosen $\lambda$ provides an optimal multiple testing
procedure in the sense of  minimizing a measure of false
non-discoveries among all rules that control a measure of false
discoveries at a specified level. These measures of false
discoveries and false non-discoveries are defined for any multiple
testing rule \mbox{\boldmath{$\delta$}} as
\begin {eqnarray}\label{def:mfdrstar}
{\rm mFDR}^*(\mbox{\boldmath{$\delta$}}) = \frac{ E \left [
\sum_{i=1}^m \delta_i (\mathbf{X}) (1 - \theta_i) w^*(\theta_i,
\mu_i) \right ] }{E  \left [\sum_{i=1}^m\delta_i (\mathbf{X}) w^*
(\theta_i, \mu_i) \right ] },
\end {eqnarray}
and
\begin {eqnarray}\label{def:mfnrstar}
{\rm mFNR}^*(\mbox{\boldmath{$\delta$}}) = \frac{ E \left [
\sum_{i=1}^m \{1-\delta_i(\mathbf{X}) \} \theta_i w^*(\theta_i,
\mu_i ) \right ]}{ E  \left [\sum_{i=1}^m \left \{1 -
\delta_i(\mathbf{X}) \right \} w^*(\theta_i, \mu_i) \right] },
\end {eqnarray}
respectively, where
\begin {eqnarray} w^*(\theta, \mu) = \left \{
\begin{array}{ll}
1 & \quad \mbox{if} \quad \theta = 0 \\
s(\mu) & \quad \mbox{if} \quad \theta = 1. \\
\end{array}
\right. \nonumber \end {eqnarray}

With $w^*(\theta_i, \mu_i)$ representing a weight associated with the $i$th hypothesis, these
measures of false discoveries and false non-discoveries can be
referred to as weighted mFDR and weighted mFNR, respectively. When
$w^*(\theta, \mu)\equiv 1$, they reduce to the corresponding mFDR
or mFNR.

\begin{theorem} \label{oracle_thm1}
Consider the model in (\ref{model}). Suppose there exists a testing procedure
$\mbox{\boldmath{$\delta$}}_0 ({\bf X})= (\delta_{10}({\bf X}),
\ldots, \delta_{m0}({\bf X}))$ such that
$\delta_{i0}({\bf X})$ is defined as in (\ref{Bayesrule}) and
$m\textrm{FDR}^*(\mbox{\boldmath{$\delta$}}_0)=\alpha$. Let
$\mbox{\boldmath{$\delta$}}({\bf X})$ be any other rule such that
$m\textrm{FDR}^*(\mbox{\boldmath{$\delta$}}) \leq \alpha$. Then
$m\textrm{FNR}^*(\mbox{\boldmath{$\delta$}}_0) \leq
m\textrm{FNR}^*(\mbox{\boldmath{$\delta$}})$.
\end{theorem}

\textbf{Proof.} First note that
\begin{align} \label{cond1}
\sum_{i=1}^m E \left [ \left \{
\delta_{i0}(\mathbf{X})-\delta_{i}(\mathbf{X}) \right \} \left \{ P(\theta_{i}=0 \mid
\mathbf{X}) - \dfrac{w_{i}(\mathbf{X})}{\lambda} P(\theta_{i}=1 \mid
\mathbf{X}) \right \} \right ] \leq 0,
\end{align} according to
(\ref{Bayesrule}), and
\begin{align}\label{cond2}
\sum_{i=1}^m E \left [
 \left \{ \delta_{i0}(\mathbf{X})-\delta_{i}(\mathbf{X}) \right \} \left \{ P(\theta_{i}=0 \mid
\mathbf{X}) - \dfrac{\alpha}{1-\alpha}w_{i}(\mathbf{X})
P(\theta_{i}=1 \mid \mathbf{X}) \right \} \right ] \geq 0,
\end{align} from the assumption, $m\textrm{FDR}^*(\mbox{\boldmath{$\delta$}}) \leq \alpha = m\textrm{FDR}^*(\mbox{\boldmath{$\delta$}}_0)$.
From (\ref{cond1}) and (\ref{cond2}), we get
\begin{align*}
\sum_{i=1}^m E\left[ \left \{
\delta_{i0}(\mathbf{X})-\delta_{i}(\mathbf{X}) \right \}w_{i}(\mathbf{X})
P(\theta_{i}=1 \mid \mathbf{X})\left(\dfrac{1}{\lambda} -
\dfrac{\alpha}{1-\alpha}\right)\right] \geq 0,
\end{align*}
which implies that
\begin{align}\label{cond3}
\sum_{i=1}^m E \left [\delta_{i0}(\mathbf{X})w_{i}(\mathbf{X}) P(\theta_{i}=1 \mid
\mathbf{X}) \right ] \geq \sum_{i=1}^m E \left [\delta_{i}(\mathbf{X})w_{i}(\mathbf{X})
P(\theta_{i}=1 \mid \mathbf{X}) \right ], \end{align}
since \[ \dfrac{\alpha}{1-\alpha} = \dfrac{\sum_{i=1}^m E \left [\delta_{i0}(\mathbf{X})
P(\theta_{i}=0 \mid \mathbf{X}) \right ]} {\sum_{i=1}^m E \left [\delta_{i0}(\mathbf{X})
w_{i}(\mathbf{X})P(\theta_{i}=1 \mid
\mathbf{X}) \right ]} \leq \dfrac{1}{\lambda}. \]
Thus, we have from (\ref{cond3})
\begin{align*}
&E\left[ \sum_{i=1}^m \left \{
\dfrac{1-\delta_{i0}(\mathbf{X})}{\sum_{i=1}^m E \left [\left \{1-\delta_{i0}(\mathbf{X})\right \}w_{i}(\mathbf{X})
P(\theta_{i}=1 \mid
\mathbf{X}) \right ]} \right. \right. - \\ & \qquad \left. \left. \dfrac{1-\delta_{i}(\mathbf{X})}{\sum_{i=1}^m E \left [ \left \{1-\delta_{i}(\mathbf{X}) \right \}w_{i}(\mathbf{X})
P(\theta_{i}=1 \mid\mathbf{X}) \right ]}
\right \} \left \{ P(\theta_{i}=0 \mid \mathbf{X})- \right. \right. \\ & \qquad \qquad \qquad \left. \left. \dfrac{w_{i}(\mathbf{X})}{\lambda}P(\theta_{i}=1 \mid\mathbf{X}) \right \}
\right] \geq 0.
\end{align*}
This implies that $$\frac{1- m\textrm{FNR}^*(\mbox{\boldmath{$\delta_0$}})}{m\textrm{FNR}^*(\mbox{\boldmath{$\delta_0$}})} \geq
\frac{1-m\textrm{FNR}^*(\mbox{\boldmath{$\delta$}})}{m\textrm{FNR}^*(\mbox{\boldmath{$\delta$}})},$$ that is, $m\textrm{FNR}^*(\mbox{\boldmath{$\delta_0$}}) \leq m\textrm{FNR}^*(\mbox{\boldmath{$\delta$}})$, as desired. \qed

\begin {remark}\label{remark:2.1} \rm %Assuming the existence of $\mbox{\boldmath{$\delta$}}_0 ({\bf X})$,
Theorem \ref{oracle_thm1} improves the work of \citet{Sun:Cai:2007} in the following sense:
1) it accommodates situations where penalties or weights
associated with type II errors can be assessed through a severity
function and incorporated into the development of a multiple testing
procedure; 2) it provides a rule that is optimal among all procedures controlling the $m$FDR* at level $\alpha$ without any distributional
restriction on the corresponding test statistics. Next, we will prove the existence of such a procedure $\mbox{\boldmath{$\delta$}}_0 ({\bf X})$.
\end {remark}

We can express the
optimal procedure $\mbox{\boldmath{$\delta$}}_0 ({\bf X})$ in
Theorem \ref{oracle_thm1} in terms of the following test statistics:
\begin{align}\label{Glfdr}
 T_{i}(\mathbf{X})=\dfrac{P(\theta_{i}=0 \mid\mathbf{X})}{P(\theta_{i}=0 \mid \mathbf{X})+ w_{i}(\mathbf{X})P(\theta_{i}=1 \mid \mathbf{X})}, i=1,\ldots,m.
\end{align}
The statistic $T_i$ will be referred to as generalized local fdr
(Glfdr). It reduces to the usual definition of
the local fdr (Lfdr) of \citet{Efron:2004} under independence and to
the test statistic defined in Sun and Cai (2009) under arbitrary
dependence when $s(\mu)=1$. We consider decision rules of the form
$\mbox{\boldmath{$\delta$}}(\mathbf{T}, c)= (\delta(T_1, c), \ldots,
\delta(T_m, c))$, where
\begin{align}
 \label{rule}
   \delta(T_i, c)= \left\{
     \begin{array}{ll}
       1 &  \textrm{if} \quad T_i \le c\\
       0 &  \textrm{if} \quad T_i > c,
     \end{array}
   \right.
\end{align}
with $c$ being such that
$m\textrm{FDR}^*(\mbox{\boldmath{$\delta$}}(\mathbf{T}, c)) \le
\alpha$. This will be our proposed oracle procedure.
Before we state this oracle procedure more explicitly
in terms of the distributions of $T_i$'s, we give the following proposition asserting the existence
of such a $c$. In this paper we assume that $\mathbf{X}$ is continuous and hence   $m\textrm{FDR}^*(\mbox{\boldmath{$\delta$}}(\mathbf{T}, c))$ is continuous in $c$.

\begin{proposition} \label{proposition:2.1}
For the decision rule in (\ref{rule}) with $T_i$ defined in (\ref{Glfdr}), $m\textrm{FDR}^*(\mbox{\boldmath{$\delta$}}(\mathbf{T}, c))$ is
non-decreasing in $c$.
\end{proposition}

We will prove this proposition by making use of the following two lemmas.

\begin {lemma}\label{lemma:1} Consider the ratio of expectations $E_{H_1} \left [\delta(T,c) \right ]/E_{H_0} \left[ \delta(T,c) \right]$, for any random variable $T$ having distribution $H_1$ in the numerator and distribution $H_0$ in the denominator. It is non-decreasing (non-increasing) in $c > 0$, if $dH_1(t)/dH_0(t)$ is non-decreasing (non-increasing) in $t$. \end {lemma}

{\bf Proof.} The ratio can be expressed as the expectation, $E_{H_c^*} \varphi (T)$, of the non-decreasing function $\varphi(T) = dH_1(T)/dH_0(T)$, where $H_c^*$ is such that
\[
dH_{c}^{*}(t) = \delta(t,c) dH_0(t)/ E_{H_0} \left [\delta(T,c) \right ].
\]
Since $\delta(t,c)$ is totally positive of order two (TP$_2$) in $(t, c)$, that is, it satisfies the inequality
\[
\delta(t, c)\; \delta (t^{\prime}, c^{\prime}) \ge \delta (t, c^{\prime}) \; \delta(t^{\prime}, c), \forall t < t^{\prime}, \; c < c^{\prime},
\]
the lemma follows from the following result \citep{Karlin:Rinott:1980}: The expectation of a non-decreasing (non-increasing) function of a random variable $Y \sim g(y, \theta)$, with  $g(y,\theta)$ being TP$_2$ in $(y, \theta)$, is non-decreasing (non-increasing) in $\theta$.
\qed

\begin {remark}\label{remark:2.2} \rm\citet{Sun:Cai:2007} derived the above result for the collection of decisions based on the test statistics satisfying the MLR condition. Note that our proof, which is different, does not rely on any such condition.
%Our essential goal is to establish the existence of the testing procedure $\mbox{\boldmath{$\delta$}}(\mathbf{T}, c)$ which is optimal in the above mentioned sense.
\end {remark}

\begin {lemma}\label{lemma:2} Given two distributions $f_0({\mathbf{x}})$ and $f_1({\mathbf{x}})$ of a random vector ${\mathbf{X}}$, define $T({\mathbf{X}}) = af_0({\mathbf{X}})/ \{ af_0({\mathbf{X}}) + b f_{1}({\mathbf{X}}) \}$, for any constants $a, b >0$. Let $H_i(t) = P_{f_i}(T({\mathbf{X}}) \le t), \; 0 < t <1$, for $i=0, 1$. Then, $dH_1(t)/ dH_0(t) = a(1-t)/bt$. \end {lemma}

{\bf Proof.} Since
\[
\left [ T ({\mathbf{X}}) - t \right ] \left [ I(T({\mathbf{X}}) \le  t ) - I(T ({\mathbf{X}}) \le t \pm \epsilon ) \right ] \le 0, \forall 0 <t <1, \epsilon >0,
\]
by taking expectations of both sides in this inequality with respect to
\[
\mathbf{X} \sim \frac{a}{a+b} f_0({\mathbf{x}}) + \frac{b}{a+b} f_1({\mathbf{x}}),
\]
we have
\[
a(1-t) \left [ H_{0}(t) - H_{0}(t \pm \epsilon ) \right ] \le  bt \left [ H_{1} (t) -H _{1} (t \pm \epsilon) \right ], \forall  0 <t <1, \epsilon > 0.
\]
The desired result then follows by letting $ \epsilon \rightarrow 0$.
\qed
\vspace{.15in}

\textbf{Proof of Proposition \ref{proposition:2.1}.}
Let $G_{i, \mbox{\boldmath{$\mu$}}}^{(j)}$ denote the conditional distribution of $T_i(\mathbf{X})$  given $\theta_i=j$ and $\mbox{\boldmath{$\mu$}}$, for $j=0,1.$  Then, from (\ref{def:mfdrstar}), we note that
\begin{eqnarray*}   m\textrm{FDR}^*(\mbox{\boldmath{$\delta$}}(\mathbf{T}, c))  =  \frac {\pi_0 \sum_{i=1}^m G_{i,0} (c) }{\pi_0 \sum_{i=1}^m G_{i,0} (c) + (1- \pi_0) \sum_{i=1}^m G_{i,1} (c) },
\end {eqnarray*}
where
\begin {eqnarray*}  G_{i,0}(c) & = & \int G_{i, \mbox{\boldmath{$\mu$}}}^{(0)}(c) h(\mbox{\boldmath{$\mu$}}|\theta_i=0) d\mbox{\boldmath{$\mu$}} \\  \mbox{and}  \qquad G_{i,1}(c) &
= & \int s(\mu_i) G_{i, \mbox{\boldmath{$\mu$}}}^{(1)}(c) h(\mbox{\boldmath{$\mu$}}|\theta_i=1) d\mbox{\boldmath{$\mu$}},
\end {eqnarray*}
with $h(\mbox{\boldmath{$\mu$}}|\theta_i=0)$ and $h(\mbox{\boldmath{$\mu$}}|\theta_i=1)$ representing the joint distribution of $\mbox{\boldmath{$\mu$}}$ conditionally given $\theta_i=0$ and $\theta_i=1$, respectively.
\begin{eqnarray} \label{proposition}
\frac{1 - m\textrm{FDR}^*(\mbox{\boldmath{$\delta$}}(\mathbf{T}, c))}
{m\textrm{FDR}^*(\mbox{\boldmath{$\delta$}}(\mathbf{T}, c))} & = &
\frac {E \left [ \sum_{i=1}^m \delta ( T_i, c) \theta_i
\omega^*(\theta_i, \mu_i) \right ] }{E \left [ \sum_{i=1}^m \delta (T_i,
c) (1 - \theta_i) \omega^*(\theta_i, \mu_i) \right ]} \nonumber
\\ & = & \frac {E \left [ \sum_{i=1}^m \delta ( T_i, c) s(\mu_i)
I(\theta_i =1) \right ] }{E \left [ \sum_{i=1}^m \delta (T_i, c)
I(\theta_i = 0) \right ]}  \nonumber \\ & = & \frac{1-\pi_0}{\pi_0}
\left (\frac{1}{m} \sum_{i=1}^m \beta_i \right ) \frac {E_{G_1}
\left [ \delta(T,c) \right]}{E_{G_0} \left [ \delta(T,c) \right] },
\end {eqnarray} where $ G_1(t) = \sum_{i=1}^m w_i \tilde G_{i,1}
(t)$, $G_0(t) = \frac{1}{m} \sum_{i=1}^m G_{i,0} (t)$, $\tilde
G_{i,1} (t) = G_{i,1}(t)/\beta_i$, and  $w_i = \beta_i/ \sum_{j=1}^m
\beta_j$, with $\beta_i = \int s(\mu_i)
h(\mbox{\boldmath{$\mu$}}|\theta_i=1) d \mbox{\boldmath{$\mu$}}$.
The proposition will be proved from Lemma \ref{lemma:1} if we can
show that $dG_1(t)/dG_0(t)$ is a non-increasing function of $t$, since the left hand side of proposition (\ref{proposition}) is a decreasing function of $m\textrm{FDR}^*(\mbox{\boldmath{$\delta$}}(\mathbf{T}, c))$.

Since $T_i(\mathbf{X}) = \pi_0 f_{i,0}(\mathbf{X})/ \{ \pi_0 f_{i,0}(\mathbf{X}) + (1 - \pi_0) \beta_i f_{i,1}^* (\mathbf{X}) \}$, and $G_{i,0}$ and $\tilde G_{i,1}$ are the cdf's of $T_i({\mathbf{X}})$ under the distributions $f_{i,0} (\mathbf{x})= f({\mathbf{x}} \mid \theta_i=0)$ and \[ f_{i,1}^*(\mathbf{x}) = \frac {1}{\beta_i} \int s(\mu_i) f(\mathbf{x} \mid \theta_i =1, \mbox{\boldmath{$\mu$}})h(\mbox{\boldmath{$\mu$}}|\theta_i=1) d \mbox{\boldmath{$\mu$}}, \] respectively, we see from Lemma \ref{lemma:2} that
$d \tilde G_{i,1}(t) = \frac{\pi_0}{(1-\pi_0)\beta_i} \left (\frac{1}{t} -1 \right ) dG_{i,0}(t)$, for any $ 0 < t <1$.  Thus,
\begin{eqnarray*}
&&\left ( \sum_{i=1}^m \beta_i \right ) dG_1(t) = \sum_{i=1}^m \beta_i d \tilde G_{i,1}(t)\\
& = & \sum_{i=1}^m \frac {\beta_i \pi_0 (1-t)}{\beta_i(1-\pi_0)t} dG_{i,0}(t) = \frac{m \pi_0}{1- \pi_0} \left ( \frac{1}{t}-1 \right ) dG_0(t),
\end{eqnarray*}
implying that $dG_1(t)/dG_0(t)$ is non-increasing in $t \in (0,1)$, as desired. Thus, the proposition is proved. \qed

%\begin {remark} \rm It is important to note from our proof of the proposition that no special distributional assumption such as monotone likelihood condition is actually required for the ultimate optimal procedure $\boldsymbol{\delta}(\mathbf{T}, c)$ with $m\textrm{FDR}^*(\boldsymbol{\delta}(\mathbf{T}, c))=\alpha$. Further, this procedure is proved to be optimal in the class of decisions rules with $m$FDR$^*$ controlled at level $\alpha$, as shown in Theorem \ref{oracle_thm1}. \end{remark}

%When $s(\mu)=1$, Proposition \ref{proposition} gives the same result that \citet{Sun:Cai:2007}  obtained by making some special distributional assumptions including a monotone likelihood condition on the test statistics. However, i
%Thus, this proposition is an improvement of the work of \citet{Sun:Cai:2007}.

\vskip 6pt Given Proposition \ref{proposition:2.1}, we are now ready to define our oracle procedure in the following:
%Let $G_{i, \mbox{\boldmath{$\mu$}}}^{(j)}$ denote the conditional distribution of $T_i(\mathbf{X})$  given $\theta_i=j$ and $\mbox{\boldmath{$\mu$}}$, for $j=0,1.$  Then, from (6), we note that \begin{eqnarray}   m\textrm{FDR}^*(\mbox{\boldmath{$\delta$}}(\mathbf{T}, c))  =  \frac {\pi_0 \sum_{i=1}^m G_{i,0} (c) }{\pi_0 \sum_{i=1}^m G_{i,0} (c) + (1- \pi_0) \sum_{i=1}^m G_{i,1} (c) } \end {eqnarray} where \begin {eqnarray} G_{i,0}(c) =  \int G_{i, \mbox{\boldmath{$\mu$}}}^{(0)}(c) h(\mbox{\boldmath{$\mu$}}|\theta_i=0) d\mbox{\boldmath{$\mu$}} \quad \mbox{and} \quad  G_{i,1}(c) = \int s(\mu_i) G_{i, \mbox{\boldmath{$\mu$}}}^{(1)}(c) h(\mbox{\boldmath{$\mu$}}|\theta_i=1) d\mbox{\boldmath{$\mu$}}, \end {eqnarray} with $h(\mbox{\boldmath{$\mu$}}|\theta_i=0)$ and $h(\mbox{\boldmath{$\mu$}}|\theta_i=1)$ representing the joint distribution of $\mbox{\boldmath{$\mu$}}$ conditionally given $\theta_i=0$ and $\theta_i=1$, respectively.

\begin{definition}[The Oracle Procedure]\label{oracle} Consider the multiple testing procedure $\mbox{\boldmath{$\delta$}}(\mathbf{T}, c^*)$, where
\begin{align} \label{oracle}
  c^*= \sup \left\{ t: m\textrm{FDR}^*(\mbox{\boldmath{$\delta$}}(\mathbf{T}, t)) \le \alpha \right\}.
\end{align}
\end{definition}

This is a generalized version of the oracle procedure of
\citet{Sun:Cai:2007}. It is developed not only under any dependence
structure among ($\mathbf{X}$, $\mbox{\boldmath{$\mu$}}$)
but also it allows the alternatives to vary across tests and each type
II error to be weighted by a measure of severity. Moreover, for its
optimality, any specific property, like the monotone likelihood
ratio property that \citet{Sun:Cai:2007} assumed, for the underlying
test statistics is not required.

%Now that we have given in the above section an oracle optimal multiple testing procedure with arbitrary dependence structure, a data-driven procedure, which can potentially be also optimal, can be developed by replacing the unknown quantities in the oracle procedure by their suitable estimates obtained from the data. More specifically, Let $\widehat{fdr}_i(\vX)$, $\widehat{w}_i(\vX)$, and $\widehat{T_i}(\vX)$ be some estimates of $fdr_i(\vX) = P(\theta_i=0|\vX)$, $w_i(\vX)$, and $T_i(\vX)$, respectively, obtained from the data. Then,
\begin {remark}\label{remark:2.3} \rm  Let $fdr_i({\bf X}) = P(\theta_i=0|{\bf X})$ and $d_i({\bf X}) = fdr_i({\bf X})/T_i({\bf X})$. Then, it is to be noted that the $m\textrm{FDR}^*(\mbox{\boldmath{$\delta$}}(\mathbf{T}, t))$ can be expressed as follows:
\begin{eqnarray*} & & \dfrac{ \sum_{i=1}^m E \left [ I(T_i({\bf X}) < t)fdr_i({\bf X}) \right ]}{\sum_{i=1}^m E \left [ I(T_i({\bf X}) < t)fdr_i({\bf X}) + I(T_i({\bf X}) < t)(1-fdr_i({\bf X}))w_i({\bf X}) \right ] } \nonumber \\ & = & \dfrac{ \sum_{i=1}^m E \left [ I(T_i({\bf X}) < t) T_i({\bf X}) d_i({\bf X}) \right ]}{\sum_{i=1}^m E \left [ I(T_i({\bf X}) < t) d_i({\bf X}) \right ] }. \end{eqnarray*}
%the above oracle procedure can be approximated (for large $m$) to the following:

%Let ${T}_{(1)}, \ldots, {T}_{(m)}$ be the ordered versions of ${T}_1({\bf X}), \ldots, {T}_m({\bf X})$. Let $H_{(i)}$ and ${d}_{(i)}({\bf X})$ be respectively the null hypothesis and the $d$-value corresponding to ${T}_{(i)}({\bf X})$. Find \begin{align} \label{datadriven}  k=\max\left\{j: \dfrac{ \sum_{i=1}^j  {T}_{(i)}({\bf X}) {d}_{(i)}({\bf X}) }{\sum_{i=1}^j {d}_{(i)}({\bf X}) } \leq \alpha \right \}. \end{align} Reject $H_{(i)}$ for all $i=1, \ldots, k$.

%A data-driven version of the oracle procedure can potentially be developed from this approximate procedure by estimating the unknown quantities in it based on the data. Of course, before using this approximate oracle procedure or its data-driven version, it would be worthwhile to establish its asymptotic equivalence to the original oracle procedure to justify its use. We will however focus on these aspects of the oracle procedure in a different %communication.
\end{remark}

\section{Numerical Studies Related to the Oracle Procedure}\label{sec:comp:oracle}
We carried out two numerical studies to see how our procedure in its oracle form compares with its relevant competitors for the problem of testing $\mu_i=0$ against $\mu_i \neq 0$, $i=1, \ldots, m$, with $s(\mu) = \mu^2$, under the following model. Let $(X_i, \mu_i, \theta_i), \; i=1, \ldots, m$, be such that
\begin{eqnarray} \label{sim:model}
\begin{array}{rcl}
X_{i} \mid \mu_{i}, \theta_i   & \ind & N(\mu_{i}, 1)  \\
\mu_{i} \mid \theta_{i} & \ind & (1-\theta_{i}) I(\mu_i=0) +\theta_{i} h(\mu_i) \\
\theta_{i} & \iid & Bernoulli(1-\pi_{0}).
\end{array}
\end{eqnarray}
%%with $\pi_0 + \pi_1 =1$, $\pi_{11} + \pi_{12} =1$.

Often a multiple testing procedure can be seen as first ranking the hypotheses according to a measure of significance, based on some test statistic, $p$-value, or local fdr, before choosing a cut-off point for the significance measure to determine which hypotheses are to be declared significant subject to control over a certain error rate, such as FDR or mFDR, at a specified level. Such ranking plays an important role in a procedure's performance, and can itself be used as a basis to compare with another procedure controlling a different error rate. More specifically, between two procedures providing the same number of discoveries, the one with better ranking should provide more true discoveries. The first numerical study was designed to make such ranking comparison between the \cite{Sun:Cai:2007} and our oracle procedures that control two different measures of false discoveries, even though one is a generalized version of the other.

Towards understanding what significance measure is being used to rank the hypotheses in our procedure, we note that under the independence model (3.1), the $m$FDR$^*(\boldsymbol{\delta}(\mathbf{T}, t))$ given in Remark 2.3 reduces to the following:
\begin{align*}
mFDR^*(\delta(\mathbf{T}, t))&=\dfrac{ E\left( I(T(\mathbf{X}) \le t)T(\mathbf{X}) d(\mathbf{X})\right)}{E\left(I(T(\mathbf{X}) \le t)d(\mathbf{X})\right)},
\end{align*}
with $T(\mathbf{X}) \equiv T_1(\mathbf{X})$ and $d(\mathbf{X}) \equiv d_1(\mathbf{X})$. The numerator and denominator expectations in the above ratio can be approximated (for large $m$) by
%the corresponding cumulative averages. That is, we can estimate $E\left( I(T(\mathbf{X}) \le t)T(\mathbf{X}) d(\mathbf{X})\right)$ and $E\left(I(T(\mathbf{X}) \le t)d(\mathbf{X})\right)$ by
$\dfrac{1}{m} \sum_{i=1}^m \left( I(T_i(\mathbf{X}) \le t)T_i(\mathbf{X}) d_i(\mathbf{X})\right)$ and $\dfrac{1}{m} \sum_{i=1}^m \left( I(T_i(\mathbf{X}) \le t) d_i(\mathbf{X})\right)$, respectively, resulting in a measure of  $mFDR^*(\delta(\mathbf{T}, t))$ at $t$ as follows:
\begin{align*}
mFDR^*(\delta(\mathbf{T}, t))&=\dfrac{ \sum_{i=1}^m I(T_i(\mathbf{X}) \le t)T_i(\mathbf{X}) d_i(\mathbf{X})}{ \sum_{i=1}^m I(T_i(\mathbf{X}) \le t)d_i (\mathbf{X})}.
\end{align*}
Let ${T}_{(1)}, \ldots, {T}_{(m)}$ be the ordered versions of ${T}_1({\bf X}), \ldots, {T}_m({\bf X})$, and  $H_{(i)}$ and ${d}_{(i)}({\bf X})$ be respectively the null hypothesis and the $d$-value corresponding to ${T}_{(i)}({\bf X})$.
Then, our oracle procedure can be described approximately as follows:

Find \begin{align} \label{datadriven}  k=\max\left\{j: \dfrac{ \sum_{i=1}^j  {T}_{(i)}({\bf X}) {d}_{(i)}({\bf X}) }{\sum_{i=1}^j {d}_{(i)}({\bf X}) } \leq \alpha \right \}, \end{align} and reject $H_{(i)}$ for all $i=1, \ldots, k$.

In other words, our procedure can be seen as ranking the hypotheses according to the increasing values of $T_{i}({\bf X})$, the Glfdr scores corresponding to the $H_i$'s, before determining the cut-off point $t \in \{ {T}_{(1)}({\bf X}), \ldots, {T}_{(m)}({\bf X}) \} $ to control the mFDR*; whereas, the Sun-Cai oracle procedure does the same in terms of the lfdr scores.

The second numerical study was conducted to see how well our oracle procedure with the cut-off point chosen subject to controlling the mFDR* compares with Sun-Cai's oracle procedure and the $p$-value based oracle procedure in \citet{Genovese:Wasserman:2002} in terms of the acceptance region, the mFDR*, and the mFNR*.

\subsection{Numerical Study 1}\label{subsec:study1}
We considered using a measure of non-discoveries to compare the rankings provided by the Sun-Cai and our oracle procedures.  More specifically, we wanted to see how these procedures compare in terms of not discovering the {\it most important} signals (i.e., the signals that are truly and highly significant), given the same number of discoveries made by each of them. The measure of non-discoveries is defined with weights assigned to the signals according to their magnitudes using our chosen severity function $s(\mu)=\mu^2$ to capture these {\it most important} signals with greater certainty.

\begin{figure} 
  \centering
  \includegraphics[width=50mm, height=50mm]{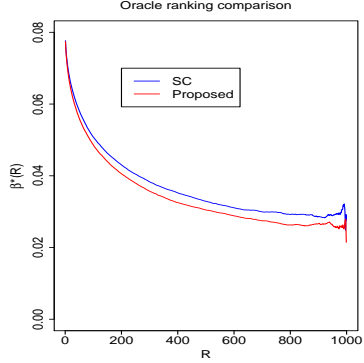}
  \caption{Simulated average weighted type II errors.}\label{fig:ranking}
\end{figure}

%given the same number of rejected null hypotheses by themdetecting the true high signal given above oracle procedure aims at controlling the mFDR* which is a generalized version of, but different from, the mFDR that is controlled by the Sun-Cai method, a meaningful . This study was designed to make a meaningful comparison between the above oracle Since the above oracle procedure aims at controlling an error rate different from that This study was designed to When testing multiple null hypotheses against their respective alternatives, a procedure can be improved by changing the underlying method of ranking the hypothesis, or considering a different error rate, or doing both. In this study, we wanted to see if the ranking based on the Glfdr is indeed better than that based on the lfdr in terms of discovering signals.
%In this work, we rank the hypotheses according to the generalized local fdr $T_i(\boldsymbol{X})$ before controlling a weighted version of the marginal FDR. \cite{Sun:Cai:2007}, on the other hand, rank the hypothesis using the local fdr, which is $T_i(\boldsymbol{X})$ with $s(\mu)=1$, before controlling the marginal FDR. %%Our  method of ranking would be better if sets of relatively low ranked null hypotheses marked for rejection contain more of those that correspond to high magnitude true alternatives, or in other words, if the null hypotheses corresponding to high magnitude true alternatives receive lower ranks by our method.
With that in mind, we generated $m=1,000$ observations according to the model (\ref{sim:model}). Here we chose $\pi_0=0.95$ and
\[
h(\mu_i) = \pi_{11}N(\mu_-, \tau^2) + \pi_{12} N(\mu_+, \tau^2),
\]
with $\pi_{11} = 0.2$, $\mu_-= -1.5$, $\mu_+=1$, and $\tau=0.5$.
We then calculated the values of Glfdr given in (\ref{Glfdr}), which can be written for this model as $Glfdr_i = \frac{\pi_0\phi(x_i)}{\pi_0\phi(x_i)+ \pi_1 H(x_i)}$ with
\begin{eqnarray*}
&&H(x_i)=\\
&=&  \pi_{11}\left[ \frac{1}{\sqrt{1+\tau^2}}\phi\left(\frac{x_i-\mu_-}{\sqrt{1+\tau^2}}\right)\frac{\tau^2}{1+\tau^2} + \frac{(\tau^2x_i+\mu_-)^2}{(1+\tau^2)} \right] \\ &+& \pi_{12}\left[ \frac{1}{\sqrt{1+\tau^2}}\phi\left(\frac{x_i-\mu_+}{\sqrt{1+\tau^2}}\right)\frac{\tau^2}{1+\tau^2} + \frac{(\tau^2x_i+\mu_-)^2}{(1+\tau^2)} \right].
\end{eqnarray*}
We ordered these values of Glfdr increasingly as $Glfdr_{(1)} \le \cdots \le Glfdr_{(m)}$. Let $H_{(i)}$ be the null hypothesis corresponding to $Glfdr_{(i)}$, for $i=1, \ldots, m$.
For each given $R=1,2,\cdots, m$, we marked the first $R$ null hypothesis to be rejected and the rest to be accepted. With $\theta_{(i)}=0$ or $1$ indicating whether the null hypothesis $H_{(i)}$ is true or false (with $\mu_{(i)}$ being the true signal), respectively, we then calculated the weighted type II errors $\sum_{j=R+1}^m \theta_{(j)} \mu_{(j)}^2$. We replicated these steps 2,000 times and averaged the 2,000 values of the weighted type II errors before obtaining the simulated value of $\beta^*(R)$, the expected weighted type II errors (or non-discoveries) given $R$ rejections (or discoveries).
The red curve in Figure \ref{fig:ranking} shows the plot of $\beta^*(R)$ against $R$.
The similar plot was obtained for the $lfdr$ score and is shown using the green curve in this figure. As seen from this figure, between the Sun-Cai and our oracle procedures, ours can potentially be more powerful in the sense of producing a smaller amount of weighted type II errors associated with missing the {\it most important} signals.

%%Out of the total 500 true alternatives, we chose only the first 150  highest magnitude ones to see where each of the corresponding null hypotheses stands in the ranking provided by the above oracle Glfdr scores. More specifically, for each $k=1, \ldots, 150$, we determined $i(k)$ such that the null hypothesis corresponding to the alternative with the magnitude $|\mu|_{(k)}$ is the one that corresponds to $Glfdr_{(i(k))}$. We repeated these steps $50,000$ times and   determined the average value of the resulting $i(k)$'s. This is the simulated value of $r(k)$, the expected rank assigned to the null hypothesis that corresponds to the $k$th highest magnitude alternative, for our method. Similarly, we determined the simulated value of $r(k)$ for the Sun-Cai method, for each $k=1, \ldots, 150$. These simulated values of $r(k)$ are plotted against $k=1, \ldots, 150$ for each of these two methods and displayed in Figure \ref{fig:ranking}. As seen from this figure, our method indeed assigns lower ranks to the null hypotheses that correspond to high magnitude alternatives.

\subsection {Numerical Study 2}\label{subsec:study2}
We chose $\pi_0=0.8$, $h(\mu_i)= \pi_{11} I(\mu_i = \mu_-) + \pi_{12} I(\mu_i = \mu_+)$ with $\mu_{-}=-3$, $\mu_{+}=4$, and let $\pi_{11}$ vary in $(0, 1)$. This model was also considered in Example 1, Section 3.2, of \citet{Sun:Cai:2007} and was chosen here to make the comparison with the \citet{Sun:Cai:2007} procedure meaningful. The rejection region for our oracle procedure is $\{X_i: X_i \leq c_{l} \; \textrm{or} \;  X_i \geq c_{u} \}$ for each $H_i$, with the cut-offs $c_l$ and $c_u$ being determined following the steps for their calculations as below:

\begin{itemize}
\item [$\bullet$]
For a given $0 < t < 1$, solve the following equation for $z$ to obtain $c_l^{(t)}$ and $c_u^{(t)}$:
 \begin{eqnarray*}
 t\pi_{1}[\pi_{11}\mu_{1}^{2}exp(\mu_{1}z-\dfrac{1}{2}\mu_{1}^2)+ \pi_{12}\mu_{2}^2exp(\mu_{2}z-\dfrac{1}{2}\mu_{2}^2)]-\pi_{0}(1-t)=0
 \end{eqnarray*} \
\item [$\bullet$]
Calculate
\begin{eqnarray*}
  && mFDR^* \\ & = & \
    \dfrac{\pi_{0}\Psi(c_{l}^{(t)},c_{u}^{(t)})}{\pi_{0} \Psi(c_{l}^{(t)}, c_{u}^{(t)}) +\pi_{1}\{\pi_{11}\mu_{1}^2 \Psi(c_{l}^{(t)}-\mu_{1}, c_{u}^{(t)}-\mu_{1}) +\pi_{12}\mu_{2}^2 \Psi(c_{l}^{(t)}-\mu_{2}, c_{u}^{(t)}-\mu_{2}) \}},
    \end{eqnarray*} where $\Psi(c_l^{(t)}, c_u^{(t)}) = 1 - \Phi (c_u^{(t)}) + \Phi (c_l^{(t)})$, and $\Phi$ is the cdf of $N(0,1)$.\
\item [$\bullet$] Repeat the above two steps until we find $t^*$ such that the $m$FDR* converges to $\alpha$.
\item $c_l$ and $c_u$ are then determined as $c_l^{(t^*)}$ and $c_u^{(t^*)}$.
\end {itemize}
Once $c_l$ and $c_u$ are determined, the mFNR$^*$ of the oracle procedure is calculated as follows:
\begin {eqnarray}
& & mFNR^* \nonumber \\
& = & \dfrac{\pi_{1}\{\pi_{11}\mu_{1}^2[1 - \Psi(c_{l}-\mu_{1}, c_{u}-\mu_{1})]+\pi_{12}\mu_{2}^2 [1 - \Psi(c_{l}-\mu_{2}, c_{u}-\mu_{2}]\}}{\pi_{0}[1 - \Psi(c_{l}, c_{u})] +\pi_{1}\{ \pi_{11}\mu_{1}^2[1 - \Psi(c_{l}-\mu_{1}, c_{u}-\mu_{1})]+\pi_{12}\mu_{2}^2[1 -\Psi(c_{l}-\mu_{2}, c_{u}-\mu_{2})]\}}. \nonumber \\
\end{eqnarray}

For the $p$-value based procedure,  the rejection region for $H_i$ is $\{X_i: |X_i| \ge c \}$ where $c$ is determined according to \citet{Genovese:Wasserman:2002}.
%% where $c$ is determined from the following equation: \begin {eqnarray} F_1(c) = \frac{\pi_0 (1 - \alpha)F_0(c)}{\alpha(1-\pi_0)}, \nonumber \end {eqnarray} with \begin {eqnarray} F_1(c) = \pi_{11} F_{\mu_1}(c) + \pi_{12} F_{\mu_2}(c), \; \mbox {and} \;  F_{\mu}(c) = 2 - \Phi (c - \mu) - \Phi (c + \mu). \nonumber \end {eqnarray} Once this $c$ is determined, the FNR of this procedure is calculated as follows: \begin {eqnarray} FNR = \frac {\pi_1 [1- F_1(c)]}{\pi_{0} [1-F_0(c)] + \pi_1 [1 - F_1(c) ]}, \nonumber \end {eqnarray} while the FNR$^*$ is determined from the equation with $c_l=-c, c_u=c$.
The oracle method of \citet{Sun:Cai:2007} is the special case of ours with $s(\mu) =1$.
%${\color{red} 1(b)}$

\begin{figure*}   
  \centering   
  \subfigure[Acceptance region]{\label{fig:accept}\includegraphics[width=50mm, height=50mm]{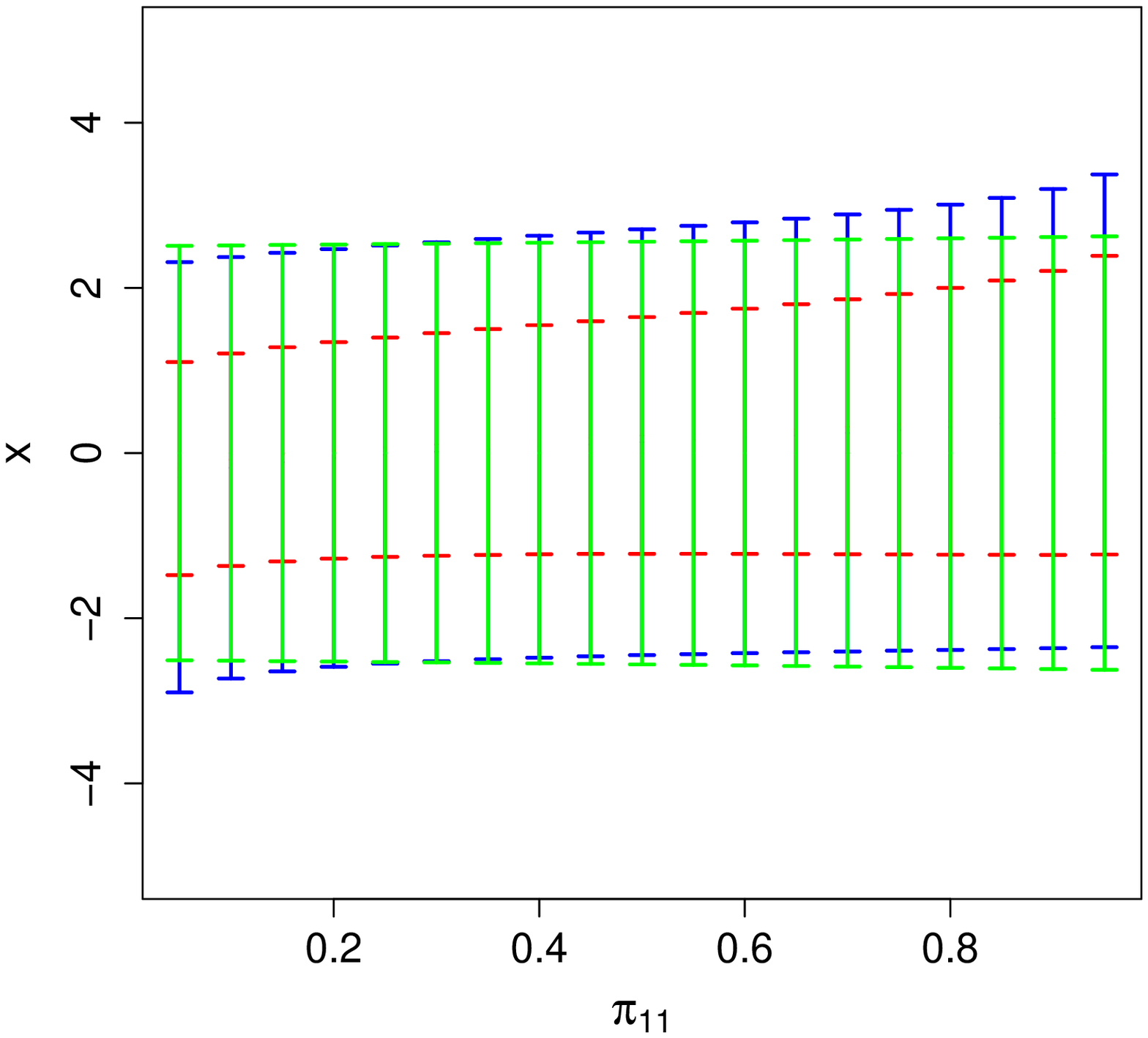}}   
  \subfigure[$m$FNR]{\label{fig:mfnr}\includegraphics[width=50mm, height=50mm]{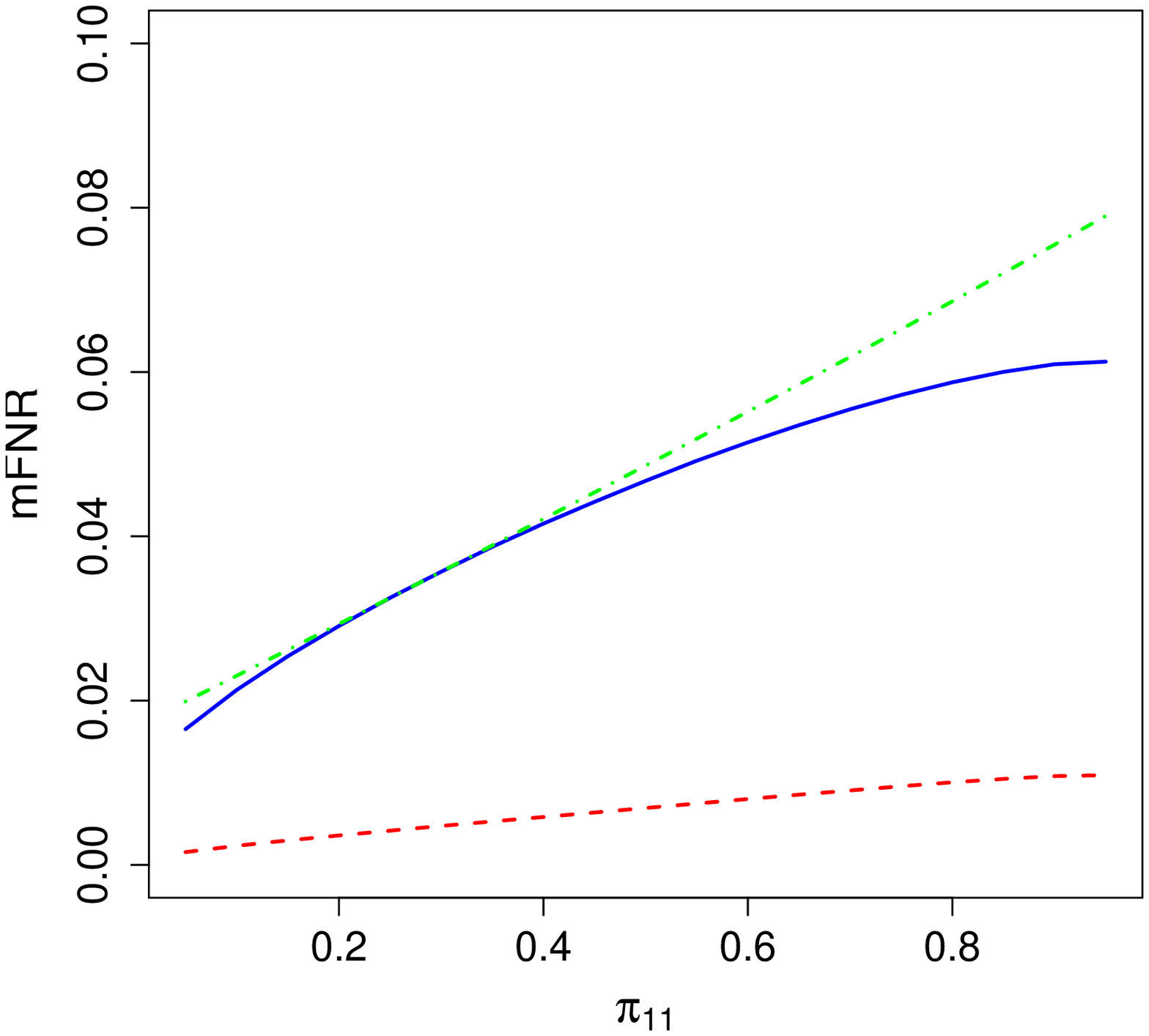}}   
  \subfigure[$m$FNR$^*$]{\label{fig:mfnrstar}\includegraphics[width=50mm, height=50mm]{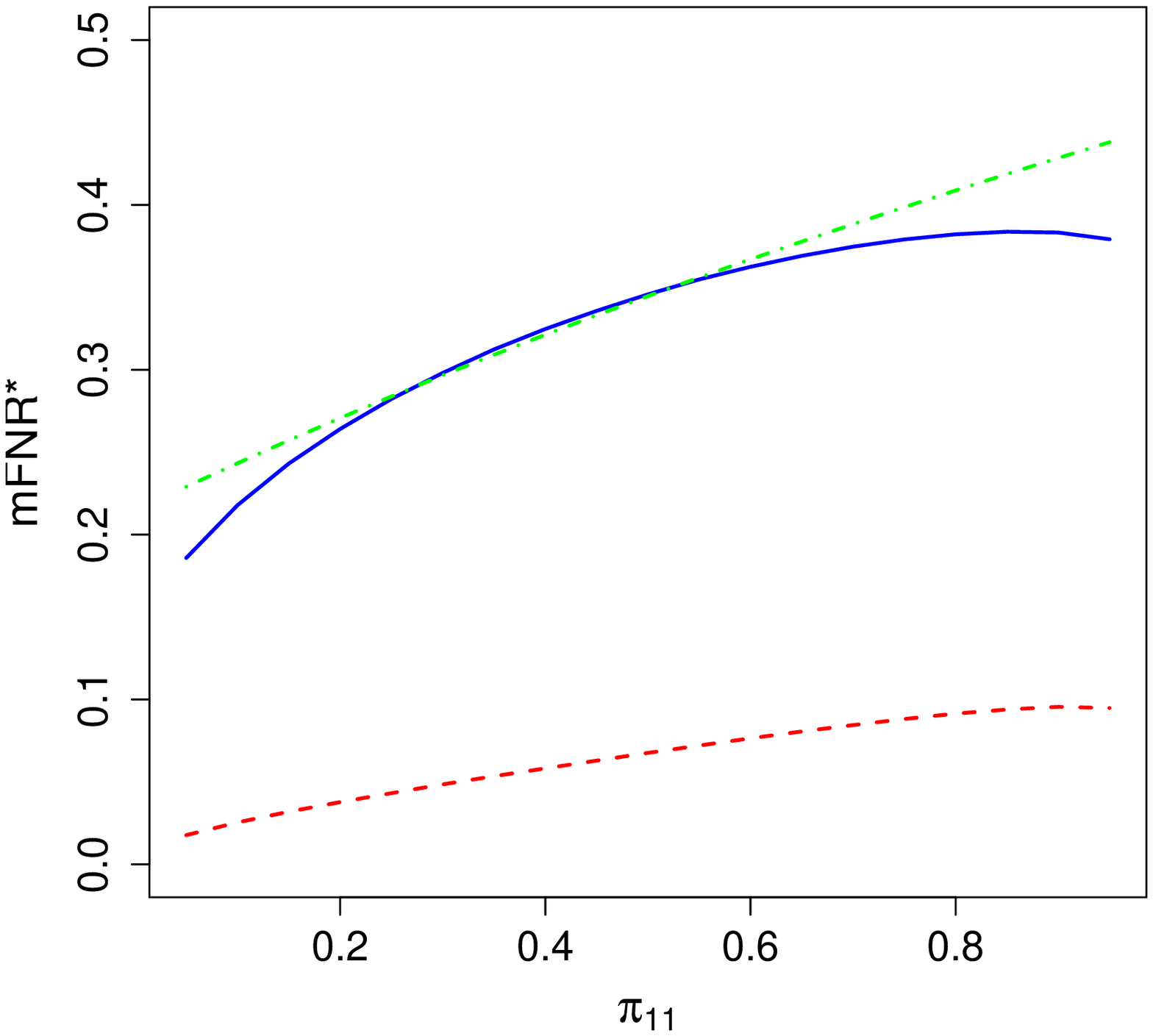}} 
  \caption{Comparison of the three procedures: (i) Our oracle procedure controlling the $m$FDR$^*$(red),   (ii) the oracle procedure of \citet{Sun:Cai:2007} controlling the $m$FDR (blue), and (iii) the $p$-value based oracle procedure of \citet{Genovese:Wasserman:2002} (green).  The data are generated according to (\ref{sim:model}) with $\pi_0=0.8$, $\pi_{11}$ varying from 0 to 1, $\mu_1 = -3$, and $\mu_2 =4$. For all three procedures, the level of control $\alpha$ is set at 0.05.}\label{fig0} 
\end{figure*}

The results of this numerical study are shown in Figure \ref{fig0}. As seen from Figure \ref{fig:accept}, the rejection regions corresponding to our oracle procedure are much wider than those corresponding to both of the other two oracle procedures. From Figures \ref{fig:mfnr} and \ref{fig:mfnrstar}, we see that while the Sun-Cai oracle procedure has smaller $m$FNR and $m$FNR$^*$ than those of the $p$-value based oracle procedure for almost all values of $\pi_{11}$, ours has the smallest $m$FNR and $m$FNR$^*$ among all three for each value of $\pi_{11}$. For instance, the ratio of the mFNR* of our procedure to that of the Sun-Cai oracle procedure can be as small as 0.15.  It is thus demonstrated that our proposed approach can potentially be more powerful than the other two approaches.

\section{Concluding Remarks}\label{sec:concluding}

The decision theoretic approach to a multiple testing problem is not
new. Other relevant work includes \citet{Sarkar:Zhou:Ghosh:2008} and
\citet{Pena:Habiger:Wu:2011}. Nevertheless, the idea of incorporating the
severity of type II errors has not been fully explored
previously in the literature. We have developed the theory behind our idea from a
compound decision theoretic point of view considering a loss function that incorporates the type II error severity. The
consideration of type II error severity into the loss function
allows us to re-formulate the work of
\citet{Sun:Cai:2007} in a more general framework involving newer,
generalized forms of marginal false discovery and marginal false non-discovery rates. Newer theoretical results generalizing and often improving the existing ones are given in this process. We now have the theory for developing a much wider class of multiple testing
procedures constructed from a decision theoretic point of view. Some
of the newer methods in this class, those corresponding to
non-constant type II error severity, are seen to have better
performance in their oracle forms, as shown in our numerical studies, than
those with constant type II error severity (i.e., those in \cite{Sun:Cai:2007} and some standard $p$-value based procedures).

The idea of weighting hypotheses or $p$-values while developing multiple testing methods in an FDR but non-decision theoretic framework has been proposed before. \cite{Benjamini:Hochberg:1997} considered weighting the hypotheses in the original definition of the FDR to define the weighted FDR and proposed a weighted version of their 1995 FDR controlling method, the so-called BH method, that controls  this weighted FDR. \cite{Genovese:Roeder:Wasserman:2006}, on the other hand, weighted each $p$-value and developed a BH type method controlling the usual FDR based on these weighted $p$-values. Our concern in this paper has been to define weighted versions of not only the marginal FDR but also the marginal FNR from their original definitions before providing a theoretical framework for the development of our procedure. Our approach to defining weighted mFDR and weighted mFNR is similar to \cite{Benjamini:Hochberg:1997}. We attach weights to the hypotheses, although they are chosen to effectively act only on the false nulls. More specifically, we have
\begin {eqnarray*}
{\rm mFDR}^*(\mbox{\boldmath{$\delta$($\mathbf{T}$,c)}}) = \frac{ E \left [
\sum_{i=1}^m I(T_i <c, \theta_i=0) \right ] }{E  \left [\sum_{i=1}^m I (T_i < c, \theta_i=0) + \sum_{i=1}^m I(T_i < c, \theta_i=1) s(\mu_i) \right ] },
\end {eqnarray*}
and
\begin {eqnarray*}
{\rm mFNR}^*(\mbox{\boldmath{$\delta$($\mathbf{T}$,c)}}) = \frac{ E \left [
\sum_{i=1}^m I (T_i > c, \theta_i=1 ) s(\mu_i) \right ]}{ E  \left [ \sum_{i=1}^m I(T_i > c, \theta_i=1) s(\mu_i) + \sum_{i=1}^m I(T_i > c , \theta_i = 0 ) \right] }.
\end {eqnarray*}
The weight is assigned to a false null hypothesis according to its signal strength. It does not depend on whether acceptance or rejection of the false null contributes to a measure of false non-discoveries or false discoveries in the form of a penalty or boon. It is important to point out that our weights for all the hypotheses don't add up to $m$, contrary to what one might conclude from \cite{Benjamini:Hochberg:1997}. In fact, a careful study of \cite{Benjamini:Hochberg:1997} would reveal that such a restriction on the weights is not necessary in their paper, even though they have assumed it.

%The weighted FDR is defined in the present decision theoretic framework as \begin {eqnarray}\label{def:weightedFDR} E \left \{ \frac {\sum_{i=1}^m \delta_i (\mathbf{X}) (1 - \theta_i) w_i^*}{ \sum_{i=1}^m \delta_i (\mathbf{X}) w_i^* + \prod_{i=1}^m [1-\delta_i (\mathbf{X})] } \right \}. \end {eqnarray}
%Our way of weighting hypotheses when defining the weighted marginal FDR is very similar to \citet{Benjamini:Hochberg:1997}. Although not considered before, the weighted FNR can be defined similarly to the weighted FDR as  \begin {eqnarray}\label{def:weightedFDR} E \left [ \frac {\sum_{i=1}^m \{1-\delta_i (\mathbf{X}) \} \theta_i w_i^*}{\sum_{i=1}^m \{1- \delta_i (\mathbf{X})\} w_i^* + \prod_{i=1}^m  \delta_i (\mathbf{X}) } \right ]. \end {eqnarray} Again, our definition of the weighted mFNR is very similar to this.

Derivation of an optimal multiple testing procedure incorporating type II error severity in its oracle form has been our primary focus in this paper. Now that we have this oracle procedure, a data-driven version of it with similar optimal property can potentially be constructed. However, construction of such an optimal data-driven procedure depends heavily on the underlying model and the chosen severity function, requiring newer efforts and techniques. We therefore leave this for a future communication. Also, a more comprehensive study of the procedure in terms of its sensitivity under varying choice of the severity function is also on our agenda for future research.

% In this paper, we obtain non-parametric empirical Bayes estimates of all the unknown quantities when assuming the severity function $s(\mu)=\mu^2$.
% For a more general form of the severity function $s(\mu)$, it is still plausible to derive the procedure by using the deconvolution of the kernel density (\cite{Stefanski:Raymond:1990}, \cite{Fan:1991}, \cite{Carroll:Hall:2004}, \cite{Delaigle:Hall:2006}) to obtain an estimate of the prior $h(\mu)$. We left it for future investigation.

\section{Acknowledgement}\label{sec:acknow}
The research of Li He is supported by Merck Research Fellowship.
Sanat K. Sarkar's research is supported by NSF Grants DMS-1006344 and DMS-1208735.
Zhigen Zhao's research is supported by NSF Grant DMS-1208735.

\end{document}